\documentstyle[11pt,aaspp]{article}
\tighten
\begin{document}
%
\def\ltsima{$\; \buildrel < \over \sim \;$}
\def\simlt{\lower.5ex\hbox{\ltsima}}
\def\gtsima{$\; \buildrel > \over \sim \;$}
\def\simgt{\lower.5ex\hbox{\gtsima}}
%
\font\smc=cmcsc10
\def\h2{H\,{\smc{II}}}
\def\s2{[S\,{\smc{II}}]}
\def\n2{[N\,{\smc{II}}]}
\def\O2{[O\,{\smc{II}}]}
\def\o3{[O\,{\smc{III}}]}
\def\Ca2{[Ca\,{\smc{II}}]}
\def\ca5{[Ca\,{\smc{V}}]}
\def\Ne3{[Ne\,{\smc{III}}]}
\def\ne5{[Ne\,{\smc{V}}]}

\title{Consequences of Dust in Metal-Rich H\,II Regions}

\author{Joseph C. Shields\altaffilmark{1} and Robert C. Kennicutt, Jr.}
\affil{Steward Observatory, University of Arizona, Tucson, AZ 85721}
\altaffiltext{1}{Hubble Fellow}

\begin{abstract}

Dust and associated depletion of heavy elements from the gas phase can
modify the thermal properties of \h2 regions from the dust-free case,
with significant consequences for the emergent optical spectrum.  We
present the results of theoretical calculations illustrating the
effects of grains on the spectra of giant, extragalactic \h2 regions,
with emphasis on high metallicity systems ($Z \simgt $Z$_\odot$).
Dust provides a simple explanation for the observational absence of
pure Balmer-line spectra that are expected on theoretical grounds for
dust-free, chemically enriched nebulae.  Grains may also play a role
in enhancements of forbidden-line emission observed in \h2 regions in
the enriched nuclei of normal galaxies.  In most cases, depletion
introduces the strongest perturbations to the optical spectrum.
Selective absorption of the ionizing continuum as well as heating by
grain photoelectrons are important in some instances, however, and
grain heating can be particularly important for enhancing emission in
high-ionization lines.  Allowing for depletion, the presence of dust
is unlikely to introduce large errors in global metallicity
indicators, although uncertainties in depletion factors coupled with
the sensitivity of infrared cooling to electron density will make
accurate calibrations difficult at high $Z$.  The present calculations
establish further that previous relative abundance analyses that fail
to take into account dust effects in a self-consistent way (grain
heating as well as depletion) may overestimate temperature gradients
in high-$Z$ nebulae, resulting in errors in relative abundances for
different elements.

\end{abstract}

\keywords{ISM: \h2 Regions -- ISM: Dust -- ISM: Abundances}

\section{Introduction}

Giant \h2 regions have a fundamental role as probes of elemental abundances in
extragalactic environments.  Existing spectroscopic studies of extragalactic
\h2 regions tend to emphasize systems with subsolar metallicities, due in part
to inherent interest in chemically unevolved systems, but also because
of selection effects, since characteristic luminosities and
emission-line equivalent widths for \h2 regions tend to decrease with
increasing source metallicity (e.g., \cite{Kennicutt88} 1988).  Data for
nebulae in relatively high-metallicity environments, such as the inner
regions of giant galaxies, are becoming increasingly available,
however, and additional theoretical work is necessary if we are to
make use of this information for quantitative abundance analyses.

The presence of abundant atomic coolants in enriched \h2 regions will
typically lead to a low equilibrium electron temperature, with a large
fraction of the radiated energy emitted by the heavy elements through
infrared ground-state fine structure transitions.  The strengths of
the comparatively energetic optical lines are very sensitive to the
electron temperature, and hence to nebular parameters that affect
temperature.  At abundances above a few times solar, the
fine-structure cooling is efficient enough that optical forbidden
lines from simple, low-density \h2 regions are expected to be
unobservable (e.g., \cite{Rubin85} 1985).  To date, however, no \h2
region emitting a pure Balmer-line spectrum has been discovered, even
in surveys of H$\alpha$-selected \h2 regions in environments for which
metallicity $Z \gg $ Z$_\odot$ is expected (\cite{OK93} 1993).  The
lack of such objects implies that physical parameters in real nebulae
are modified so that the electron temperature is high enough to
produce significant optical forbidden-line emission.

The effects of interstellar dust could conceivably underlie at least
some of the enhancement of optical emission in high-$Z$ \h2 regions.
Dust can act to modify the thermal structure of a nebula in several
ways.  The removal of atomic coolants from the gas phase via depletion
will tend to increase the electron temperature, leading to measurable
changes in the emitted spectrum (e.g., \cite{Henry93} 1993).  Grains
may selectively absorb a fraction of the Lyman continuum photons
powering the nebula, resulting in modification to the ionizing
spectral energy distribution seen by the gas and to the resultant
heating (e.g., \cite{Sarazin77} 1977; \cite{Mathis86} 1986;
\cite{Aannestad89} 1989).  Grains will also participate directly in the
thermal balance by contributing photoelectrons that heat the plasma,
and by cooling through radiation of energy transferred in captures of
electrons (e.g., \cite{Olive86} 1986; \cite{Baldwin91} 1991).

Previous studies emphasizing nebulae at $Z \simlt $ Z$_\odot$ have
generally concluded that selective absorption of the ionizing
continuum by dust has a relatively small effect on the emergent
optical spectrum of diffuse \h2 regions (e.g., \cite{Mathis86} 1986).
Given the greater temperature sensitivity of optical lines at higher
$Z$, we might expect the influence of dust to be more important in
metal-rich nebulae; this statement is especially true if the
dust-to-gas ratio scales with $Z$.  The observational consequences of
grain heating and cooling within \h2 regions over the entire span of
metallicity have also received only limited attention in previous work.

Motivated by these considerations, we have undertaken photoionization
calculations appropriate for giant extragalactic \h2 regions in order
to understand possible consequences of dust in chemically enriched
nebulae.  We find that dust and associated depletion can indeed
produce nonnegligible perturbations in the optical spectrum of
metal-rich \h2 regions.  The influence of dust in such objects
provides a possible basis for explaining several unexpected empirical
characteristics of \h2 regions in high-$Z$ environments, including the
lack of pure Balmer-line nebulae.  If grains survive in the inner
parts of spherical nebulae, they may selectively enhance emission in
high-ionization lines.  Dust is unlikely to contribute large
uncertainties in {\it gas phase} abundance estimates based on the
oxygen emission lines, but ambiguities in depletion factors and
density effects will contribute substantial uncertainty to
``empirical'' metallicity-excitation calibrations for $Z \simgt $
Z$_\odot$ (e.g., \cite{Pagel79} 1979).  Finally, we suggest possible
revisions to theoretical relations between ion-weighted temperatures
employed in relative abundance analyses, necessitated by a
self-consistent treatment of the presence of dust in \h2 regions.

\section{Nebular Calculations with Dust}  \label{Calculations}

Calculations appropriate for prediction of \h2 region optical spectra were
generated with the photoionization code Cloudy, version 84.09
(\cite{Ferland94} 1994).  Parameters were chosen such that the results would
illustrate the possible consequences of dust in giant, extragalactic \h2
regions (GEHRs) as a function of metallicity.  The ionizing continuum was
represented by \cite{Kurucz79} (1979) stellar atmospheres corresponding to
$T_\star$ of 38,000 K and 45,000 K, with total ionizing luminosities (photons
s$^{-1}$) of log $Q$(H) $= 50.7$ and 51.2, respectively.  The cooler stellar
continuum model is interpolated from theoretical calculations for $T_\star = $
35,000 K and 40,000 K, and both of the final continuum choices assume stellar
log($g$) = 4.5.  Based on systematic trends in existing observations, the
lower $T_\star$ values are more applicable for nebulae with
relatively high $Z$ (e.g., McCall, Rybski, \& Shields 1985;
\cite{Vilchez88} 1988).

The nebular geometry was assumed to be spherical, extending from an
inner radius of 0.01 pc to the Str\"omgren radius $R_s$ defined as the
point where the ionized hydrogen fraction fell below 1\%.  A total
hydrogen density of $n_H = 10$ cm$^{-3}$ was adopted for material
described by volume filling factors of $\epsilon = 0.0617$ and 0.0347
for the 38,000 K and 45,000 K cases, respectively.  If we define
ionization parameter $U$ in the customary way [$U
\equiv Q$(H)$/(4\pi R_s^2n_e c$), where $c$ is the speed of light],
these combinations of parameters yield log $U = -3.0$ at $Z = $
Z$_\odot$ in the dust-free case; estimates of $U$ for GEHRs are
comparable to or somewhat larger than this value (\cite{Shields90} 1990
and references therein).  $U$ is a function of $Z$, due to the
dependence of recombination rates and hence $R_s$ on electron
temperature $T_e$, as well as a function of dust content, since grains
will absorb a portion of the ionizing radiation and also modify the
$T_e$ structure of the nebula through heating and cooling processes.

Total elemental abundances for the case of $Z = $ Z$_\odot$ were taken from
\cite{Grev89} (1989).  For other values of $Z$, abundances of elements
heavier than helium were assumed to be in solar proportions, with the
exception of nitrogen, which was assumed to have a secondary component
such that its abundance scales as $Z^2$.  Empirical evidence for a
secondary contribution to N in GEHRs is discussed, for example, in
\cite{Vila93} (1993).  Depletion onto grains was assumed to modify
gas-phase abundances by factors taken from \cite{Cowie86} (1986, Table
3) corresponding to an overall average for the interstellar medium
(ISM).  The resulting total and depleted abundances at $Z = $ Z$_\odot$
are listed in Table 1.

The grains themselves were assumed to consist of a graphite-silicate
mix with a size distribution as described by \cite{Mathis77} (1977)
for the diffuse ISM, with optical constants as given by
\cite{Draine84} (1984) and extended within the extreme ultraviolet
(EUV) bandpass by \cite{Martin89} (1989).  The resulting dust opacity
features a maximum at photon energies of $\sim 17$ eV and a decrease at
higher energies, as required by the Kramers-Kronig relations for realistic
grain materials (e.g., \cite{Bohren83} 1983; \cite{Mathis86} 1986).

Scattering by grains is neglected in the present calculations.  At
ionizing energies, the scattering phase function for the grains is
expected to be predominantly forward-throwing.  At optical
wavelengths, scattering in and out of the line of sight will
approximately cancel in a spherical geometry.  In the current
implementation of Cloudy, reddening of the emergent spectrum by grain
absorption is also neglected in the spherical case.  Our principal
interest in the present study is the consequences of dust for the
structural and thermal characteristics of nebulae, rather than on
emission-line
transfer effects of the dust.  The effects of internal scattering and
absorption have been discussed previously by \cite{Mathis70} (1970, 1983),
\cite{Petrosian80} (1980), and \cite{Aannestad89} (1989), among others.
For the cases of interest here, the scattering and absorption optical
depths within $R_s$ are relatively small (\S\ref{extinct}); in such
cases, dereddening of the observed optical spectrum based on the
Balmer line strengths is expected to reproduce the intrinsic line
ratios with reasonable accuracy (\cite{Mathis83} 1983).  As a result,
the predictions of line ratios presented here are suitable for direct
comparison with dereddened ratios derived from observations.

Thermal effects of dust are quantified by extending the calculation of
nebular ionization and energy equilibrium to include grains.  Additional
details concerning the treatment of grain physics within Cloudy can be found
in \cite{Baldwin91} (1991, Appendix C) and \cite{Ferland94} (1994).  The
dust-to-gas ratio was assumed to scale in proportion with $Z$, with a
normalization of 0.01 by mass at $Z = $ Z$_\odot$.

Giant \h2 regions exhibit supersonic line widths that correlate in
amplitude with emission-line luminosity (\cite{Melnick79} 1979).  Line
widths in excess of thermal values will modify the transfer of lines
including infrared fine-structure transitions that would otherwise be
optically thick in a quiescent nebula.  The added velocity width
facilitates the escape of these lines, and thus may decrease the
equilibrium $T_e$ when the infrared transitions contribute a
significant fraction of the total cooling (e.g., \cite{Simpson75}
1975).  We consequently assumed the presence of a microturbulent
velocity component $b_{turb}$ added in quadrature with the thermal $b$
($\equiv [2kT_e/m]^{0.5}$), with values of $b_{turb} = 17.4$ km
s$^{-1}$ for the case of $T_\star = 38,000$ K and log $Q$(H) $= 50.7$,
and $b_{turb} = 23.0$ km s$^{-1}$ for $T_\star = 45,000$ K and log
$Q$(H) $= 51.2$.  These choices of $b_{turb}$ are derived from the
empirical relation between recombination line luminosity and velocity
width given by
\cite{Arsenault88} (1988).

\clearpage

\section{General Results} \label{general}

The predicted strengths of optical forbidden lines, normalized to
hydrogen recombination features, are plotted as a function of $Z$ in
Figures 1 and 2 for $T_\star = 38,000$ K and 45,000 K, respectively.
The curves show the expected behavior with increasing $Z$, with
forbidden-line emission initially rising in response to the growing
number of emitters, reaching a maximum, and declining at high $Z$ as
the total cooling efficiency increases and $T_e$ diminishes.  The
ratio of \O2 $\lambda$3727/H$\beta$ exhibits an additional reversal,
with increasing values at very high $Z$.  This behavior stems from
photoionization of O$^0$ into excited states of O$^+$ that can
subsequently decay radiatively.  The resulting emission component
consequently scales directly with the abundance of O and hence with
$Z$.  While such emission can be important in some contexts (e.g.,
novae; \cite{Ferland81} 1981), it is unlikely to be significant in \h2
regions, for which \O2/H$\beta$ $\simgt 0.3$ observationally (e.g.,
\cite{OK93} 1993).

The separate curves in a given plot in Figures 1 and 2 illustrate the change in
emission line behavior as the effects of dust are added incrementally.  The
solid curve corresponds to dust-free, undepleted nebulosity, while the other
curves show the results of introducing depletion, dust opacity, and grain
heating $+$ cooling. In almost all cases, the effects of dust lead to an
enhancement of optical forbidden-line emission over the undepleted, dust-free
case, due to an increase in equilibrium $T_e$.

For most of the lines, the largest shift in the plotted line ratio
curves results when depletion is introduced.  This pattern implies
that depletion of gas-phase coolants is likely to be the most
important aspect of dust in terms of its influence on the observed
spectrum (see also \cite{Mathis86} 1986); this statement is supported
by comparisons of how forbidden line strengths change when the three
effects of dust are added in isolation, rather than in succession.
While depletion of optically emitting elements can produce a
reduction of line emission at low $Z$, the result is more typically an
increase in optical line emission as the cooling efficiency of other
transitions decreases and equilibrium $T_e$ increases.  Infrared
fine-structure transitions are particularly important for establishing
this behavior, since they scale in strength with abundance but are
relatively insensitive to $T_e$. \cite{Henry93} (1993) has emphasized
the importance of the [Si\,{\smc{II}}] 34.8 $\mu$m line and infrared
[Fe\,{\smc{II}}] transitions for cooling in an undepleted nebula; in
real nebulae, these elements are likely to be subject to significant
depletion onto grains, producing an increase in $T_e$ that can boost
the strength of optical lines.

Selective absorption of ionizing radiation by dust additionally modifies the
ionization and thermal structure of a nebula, leading to changes in its
optical spectrum.  The decrease in grain opacity at photon energies
above the Lyman limit results in
a hardening of the ionizing radiation field as it is attenuated by grains.
The resultant boost in average photon energy produces greater heating per
ionization, with a corresponding increase in $T_e$ and optical line emission.
The amplitude of this effect in GEHRs is likely to be
small, based on the results in Figures 1 and 2 as well as analytic
arguments.  The optical depth of dust will scale with the total column
density of gas within the Str\"omgren radius, $N_H = n_H R_s$, and the
dust-to-gas ratio, assumed here to scale with $Z$.  From the definition of
$U$, we can express \begin{equation} N_H = U {{3c}\over{\alpha_B}} \approx 3
\times 10^{20}\left( U\over{10^{-3}}\right) {\rm cm}^{-2},\end{equation}
where the case B hydrogen recombination coefficient $\alpha_B$ for $T_e =
7500$ K (approximately correct for $Z = $ Z$_\odot$) and $n_e = 100$ cm$^{-3}$
is taken from \cite{Hummer87} (1987).
For a dust-to-gas ratio representative of the Galactic ISM, the maximum EUV
absorption optical depth will occur at approximately 17 eV, with $\tau_{EUV}
\approx (2 \times 10^{-21} {\rm cm}^2) N_H$ (\cite{Martin89} 1989).
For a Str\"omgren sphere, the maximum EUV optical depth with thus be
\begin{equation} \tau_{EUV} \approx 0.6 \left( U\over{10^{-3}}\right)
\left( Z\over{{\rm Z}_\odot}\right), \end{equation} where $U$ is defined
in the dust-free limit.  The expected $\tau_{EUV}$ in GEHRs is thus
significant for typical parameters, and will be higher if the nebulae
are relatively dusty and/or $U$ is elevated above typical values.
Substantial modification of the ionizing spectral energy distribution
is more likely in Galactic ultracompact and blister \h2 regions (e.g.,
\cite{Aannestad89} 1989), which may be described by $U$ values much
greater than those typical of GEHRs.  When dust opacity is large,
$R_s$ and $\tau_{EUV}$ are obviously coupled further by the diminution
of ionizing flux seen by atomic material, a point considered in more
detail in \S\ref{extinct}.

Figures 1 and 2 also demonstrate that grain heating boosts the predicted
optical emission even further.
Photoelectrons emitted by grains are expected to be
more energetic and hence produce greater heating than photoelectrons
emitted by the constituent atoms in isolation, because the ionization
potential, or work function, is reduced for matter in a solid phase
(e.g., see \cite{Draine89} for a review).  The treatment of grains
employed by Cloudy assumes a minimum work function of 8 eV for a
neutral grain, with an increase in threshold energy as the grain
becomes more strongly depleted of electrons and charged.  The heating
rate scales directly with the density of grains and, for a weak
grain potential, the flux of radiation above the threshold energy.  In
a dust-free nebula, heating tends to be dominated by photoelectrons
from recombined hydrogen and hence scales as $n_e^2$.  For an \h2
region with fixed density and dust-to-gas ratio, the significance of
grain heating relative to hydrogen bound-free heating consequently
scales with the flux $\phi$ of energetic radiation (or more generally, with
ionization parameter $U$, defined locally as a ratio of ionizing
photon to electron densities).  The relative dependence of grain
heating on $\phi$ introduces a radial variation in the resulting thermal
effects for a spherical nebula, with greater heating occurring at
small radii (see \S\ref{tempgrads}).  Grain cooling by electron
attachment is also likely to be significant, and the relative
importance of heating and cooling by grains is a more complicated
function of the ionizing spectral energy distribution, grain
potential, and $T_e$ (\cite{Baldwin91} 1991, Appendix C.14).

These initial results suggest that the collective effects of dust may indeed
account for the absence of observed pure Balmer-line nebulae.  While \o3
emission may become unobservable in low excitation GEHRs, \O2 emission is
invariably detected, as noted previously, with \O2 $\lambda$3727/H$\beta$
$\simgt 0.3$.  This minimum is consistent in the present calculations with $Z$
as high as $2 - 3 \times $ Z$_\odot$.  Observed strengths of \n2 and \s2
are similarly consistent in general terms with metallicities as high as this
bound.


\section{Verisimilitude}

\subsection{Grain Assumptions}

The calculations presented here make a number of assumptions of uncertain
validity for GEHRs.  The greatest uncertainty in this discussion is the
detailed nature of grains in these nebulae.  From studies of the Galactic ISM,
it is well established that depletion rates are a function of environment
(\cite{Cowie86} 1986 and references therein).
Given the substantial output of radiative and kinetic energy by young stars,
grains located near these sources might be expected to be modified and/or
partially destroyed.  \cite{Strom74} (1974) presented a comparison of infrared
emission from Galactic and extragalactic \h2 regions that has been interpreted
as evidence for grain destruction in the GEHRs.  Galactic sources exhibit a
correlation between thermal radio continuum emission and infrared luminosity
that \cite{Strom74} employed to predict infrared fluxes from a group of more
luminous GEHRs, using H$\beta$ emission in place of radio continuum emission to
trace $Q$(H).  Observed $10 - 20$ $\mu$m fluxes for the extragalactic sources
fall short of the predicted fluxes derived in this manner.  \cite{Strom74}
discussed several possible explanations for this discrepancy, including a
deficiency of dust in the luminous extragalactic sources.

The difference noted by \cite{Strom74} between measured IR fluxes and the
extrapolated predictions may be a manifestation of a more general pattern of
disparity between high- and low-luminosity sources.  As discussed by
\cite{Habing79} (1979) and
\cite{Kennicutt84} (1984), \h2 regions show systematic trends with luminosity
such that lower luminosity sources are denser and more centrally
concentrated than their giant counterparts.  Both tendencies might be
expected to result in higher $U$ for the Galactic sources, and
consequently greater heating of dust grains and resultant IR emission,
consistent with the observational trend found by \cite{Strom74}.
Alternatively, studies of the Galactic ISM demonstrate that depletion
and hence the dust-to-gas ratio correlate with density
(\cite{Mathis90} 1990), which would be consistent with a higher grain
content in dense Galactic \h2 regions when compared with more diffuse
GEHRs.

Recent clues to the grain content of GEHRs are provided by analyses of
silicon and iron abundances in \h2 regions and compact blue galaxies
by \cite{Garnett95} (1995) and Thuan (1995, private communication).
In both studies the results suggest that the depletion rates of these
elements are lower in GEHRs than in the average ISM.  These findings
support a picture in which a fraction of the grains in nebulae are
destroyed, such that the dust-to-gas ratio in GEHRs is reduced
somewhat from a simple extrapolation based on the Galactic ISM.  If
these results are characteristic of GEHRs, the nebular predictions
described here will exaggerate to some degree the actual effects of
dust in these sources.

The present calculations also assume a specific grain size
distribution, which is again known to be a function of environment.
The EUV grain opacity and rate of grain heating are very sensitive to
the small grain component; if the small grain population is reduced
compared to the ISM average, as inferred in the Orion nebula and other dense
environs (\cite{Mathis90} 1990 and references therein), the effects of
dust opacity and heating will be correspondingly diminished.  If 30
Doradus can be regarded as representative of GEHRs, however, the small
grain component may actually be {\it enhanced} relative to grains in
the Galactic ISM (Fitzpatrick 1986).  The elemental content of dust in
GEHRs is a further major uncertainty.  The relative depletion rates
are poorly constrained, and more generally, the {\it total} relative
abundances may differ from a simple scaling of solar proportions.  In
light of these many fundamental ambiguities, the present calculations
can only be considered illustrative of the effects of dust in
\h2 regions, rather than providing definitive predictions.

\subsection{Line Ratios}

One means of assessing whether the nebular calculations described here
bear any applicability to real GEHRs is by comparison of predicted and
observed line ratios for these systems.  Figure 3 shows line ratio
plots selected for ease of comparison with those given in Figure 8 of
the extensive study of GEHRs by \cite{McCall85} (1985); the
plotted points represent observational measurements from
\cite{McCall85}.  The plotted curves show the prediction
results for a dust-free nebula (solid curves) and for a depleted
nebula with full grain physics (dotted curves), as a function of
changing $Z$, for the two choices of $T_\star$.  The behavior of the
oxygen lines suggests that a higher $T_\star$ is required at lower $Z$,
in agreement with previous work (e.g., \cite{McCall85} 1985).

Figure 3 shows that in the high metallicity regime ($Z \simgt {\rm
Z}_\odot$), the calculations for the undepleted nebula tend to
underpredict emission in \n2 and \s2, while the flux in these lines is
excessive in the calculations with grains.  This pattern is more
readily seen in Figure 4, which plots \o3/H$\beta$ as a function of
\n2/H$\alpha$ and \s2/H$\alpha$ for the same data sets.  A simple
explanation for this behavior might be that the assumed grain
component is excessive, and the actual depletion and grain effects are
somewhat reduced.  Other explanations are possible, however.  The
strength of \s2 emission is particularly sensitive to the value of $U$
(e.g., \cite{Skillman89} 1989), and additional calculations indicate
that the predicted \s2/H$\alpha$ ratio for the dusty nebula can be
brought into agreement with the observations by increasing $U$ by a
factor of $\sim 2 - 3$.  This remedy is less successful for the \n2
line strength, however, and fine-tuning of other parameters such as N
abundance is likely to be necessary to bring predictions for \n2 into
agreement with the observations.  Decreasing $U$ in the dust-free case
does not significantly increase \s2, and leads to substantial
decreases in the strength of the other optical forbidden lines
considered here.

The predictions in Figure 4 permit an interesting comparison with
similar plots in Figures 7 and 8 of Kennicutt, Keel, \& Blaha (1989)
for observations of \h2 regions in the nuclei of galaxies.
\cite{KKB89} (1989) noted that the nuclear sources show a systematic
trend of greater \n2 and \s2 line strength at a given value of
\o3/H$\beta$ than is seen in extranuclear GEHRs.  Their
observational results for the nuclear sources are, in fact, in
relatively good agreement with the predictions for dusty nebulae
presented here, for $Z$ up to $\sim 2$Z$_\odot$.  \cite{KKB89}
suggested that the trend in the nuclear \h2 regions might stem from
the presence of weak active nuclei that are spatially unresolved from
normal circumnuclear star-forming regions.  The present results
suggest a possible alternative explanation whereby the enhanced \n2
and \s2 emission results from an elevated dust content in nuclear
nebulae relative to disk \h2 regions.  The nucleus corresponds to the
center of a galaxy's potential well, and the interstellar pressure in
such an environment is expected
to exceed typical disk ISM pressures by up to 2 orders of
magnitude or more (e.g., \cite{Spergel92} 1992; \cite{Helfer93} 1993).
Resulting enhancements in density may
foster enhanced grain formation and growth, consistent with local
correlations between density and depletion, thus providing a physical
motivation for why nuclear nebulae could be unusually dusty.

\subsection{Observable Extinction} \label{extinct}

If grains are important in modifying the overall behavior of an \h2
region, we might expect this fact to be signaled observationally by
signatures of reddening or extinction.  The extinction optical depth
produced by dust associated with the \h2 scales as in equation 2, but
with optical depth at $V$ ($\tau_V$) reduced by a factor of $\sim 5$
from $\tau_{EUV}$ (\cite{Martin89} 1989), such that $\tau_V$ will be
only a few tenths as the dust column becomes optically thick at
ionizing energies.

If the dust optical depth is increased beyond this level, the fact
that the grain EUV and optical opacities are coupled implies an
additional feedback mechanism that will limit optical indications of
extinction.  Significant $\tau_{EUV}$ reduces $N_H$; but achieving a
large $\tau_{EUV}$ requires a substantial $N_H$ for typical
dust-to-gas ratios.  Large EUV opacity and hence significant $\tau_V$
will thus result only from extreme nebular conditions.  For example,
if internal dust is sufficient to produce $\tau_V \approx 1$ and hence
$\tau_{EUV} \approx 5$, then $N_H$ (which is proportional to
$[Q(H)n\epsilon^2]^{1/3}$ in the dust-free case) must be $\sim 3
\times 10^{20}$ for $Z = $ Z$_\odot$ (cf. eqn. 1).  But $Q(H)$ seen by
the gas will be reduced as $\tau_{EUV}$ increases and a growing
fraction of the ionizing radiation is absorbed by grains; $\tau_{EUV}
\approx 5$ will result in a reduction of the nebular radius by a
factor of $\sim 2$ relative to the dust-free $R_s$, with $\sim 90$\%
of the ionizing photons absorbed by dust for a standard dust-to-gas
ratio (\cite{Petrosian72} 1972; see also \cite{Mathis71} 1971 and
\cite{Aannestad89} 1989).  In order to obtain the $N_H$ producing
$\tau_V \approx 1$, the product $[Q(H)n\epsilon^2]^{1/3}$ and hence
$U$ corresponding to the dust-free case must be increased by
approximately an order of magnitude.

Conditions in which dust so strongly modifies the nebula are probably
only present in systems that are clearly distinct from classical
GEHRs.  One such class of objects is that of ultracompact \h2 regions
(e.g., \cite{WC89} 1989), which are far denser than typical GEHRs and
show observational evidence for substantial dust absorption of
ionizing radiation.  Similar conditions may also obtain in the dense
central regions of starburst galaxies.  The high $n_e$ and large $U$
that describe these environments are inconsistent with optical line
ratios for GEHRs.

Substantial extinction may still be present with less dramatic
effects on the nebula if the ionization is produced by distributed
sources rather than a central compact star cluster.  An idealized
picture illustrating this point would be a collection of Str\"omgren
spheres aligned along the line of sight; $\tau_V$ through a single
sphere may be small, but the collective effect of multiple foreground
systems could produce significant extinction of emission produced in
the more distant spheres.

Observations suggest, however, that internal optical
extinction in GEHRs is indeed fairly small.  Studies of the stellar
population in 30 Doradus (\cite{Malamuth94} 1994) and in NGC~3603
(\cite{Moffat94} 1994), for example, demonstrate that the massive
stellar component that dominates production of ionizing radiation is
relatively concentrated, suggesting that idealization of the nebula as
a single Str\"omgren sphere is unlikely to be grossly in error.
Empirical estimates of $A_V$ for GEHRs are also rarely more than $A_V
\approx 1$ (\cite{Kennicutt84} 1984).  While extinction values based
on reddening of the Balmer lines alone may underestimate the total
extinction, due to surface effects (\cite{Mathis70} 1970), radio
measurements of GEHRs suggest that the optical estimates of total
$A_V$ are not in error by large factors (\cite{Israel80} 1980; \cite
{Caplan86} 1986; \cite{vanderHulst88} 1988; see also discussion in
\cite{Shields90} 1990).

Although the total optical extinction in GEHRs is thus only modest,
the effect of dust on the nebula can still be significant if EUV
opacity is much larger than opacity at optical wavelengths.  It also
must be remembered that attenuation of the {\it ionizing} radiation
field is not directly traced by the comparison between Balmer line and
radio fluxes.


\section{Further Implications}

\subsection{Global Metallicity}

Existing abundance analyses of \h2 regions have relied heavily on the
$R_{23}$ parameter [$\equiv$ (\O2 $\lambda$3727 $+$ \o3
$\lambda\lambda 4959, 5007$)/H$\beta$] as an indicator of the
abundance of oxygen, which in turn is often taken as a proxy for the
total metallicity (\cite{Pagel79} 1979).  This approach can be
justified by the dominant role of oxygen as a coolant in typical
nebulae; the calibration of $R_{23}$ as a function of $Z$ is
nonetheless dependent at some level on the abundances of other
coolants, however.  An important question is thus whether this
calibration is significantly modified by the presence of grains and
associated depletion.  Another way of expressing this question is to
ask whether the gas-phase oxygen abundance derived from $R_{23}$ can
be translated into a total oxygen abundance (or $Z$) through simply
scaling by the oxygen depletion factor.

To illustrate the effects of dust on the use of $R_{23}$ as an abundance
indicator, Figure 5 displays $R_{23}$ as a function of
oxygen abundance {\it in the gas phase}, for the calculations described
in \S\ref{Calculations}.  By comparing the curves for a given $T_\star$
that correspond to no dust and full dust effects, we can see that relatively
small changes in $R_{23}$ result from the presence of grains and
associated depletion at low $Z$.  The shift in predicted $R_{23}$ due
to grains at a given gas-phase abundance increases significantly at
higher $Z$; at the same time, however, $R_{23}$ is becoming a steeper
function of $Z$, such that a large shift in predicted $R_{23}$ propagates into
a small difference in $Z$.  In terms of the observationally relevant
comparison, for a measured value of $R_{23}$, the uncertainty in
$Z$ due to dust effects is likely to be $\simlt 0.2$ dex for a gas-phase
oxygen abundance of up to $\sim 3$ times the solar value.  We conclude
that the effects of grains are unlikely to introduce large errors in
measurements of gas-phase oxygen abundance based on $R_{23}$.  This
conclusion differs somewhat from that of \cite{Henry93} (1993).
The calibration of $R_{23}$ at high $Z$ is rather sensitive to $n_e$,
however, due to suppression of fine-structure cooling (\cite{OK93} 1993),
as well as $U$, and the cumulative effects of these uncertainties cast
doubt on whether a reliable calibration of $R_{23}$ versus $Z$ is possible
at $Z > $ Z$_\odot$ (see also \cite{Mathis86} 1986).


\subsection{Temperature Gradients} \label{tempgrads}

Temperature gradients in dust-free \h2 regions are determined by the
efficiency of cooling mechanisms and the average heating per
ionization as a function of radius.  At very low $Z$, collisionally
excited hydrogen Ly$\alpha$ plays an important role in cooling.  The
neutral fraction of hydrogen that can give rise to such emission
increases monotonically with radius, so that the cooling efficiency
increases with $r$, and $T_e$ tends to be a decreasing function of
radius.  At higher $Z$, collisionally excited lines of heavy elements
become important for cooling, with infrared transitions playing an
increasing role; important ions with ground-state fine structure
giving rise to infrared emission include doubly ionized O, S, and N,
which boosts the cooling efficiency in the inner nebula where these
species are abundant\footnote{The resulting lines are \o3 51.8 and
88.4 $\mu$m, [S\,{\smc III}] 18.7 and 33.5 $\mu$m, and [N\,{\smc III}]
57.3 $\mu$m.} (\cite{Stasinska80} 1980).  In addition, the frequency
dependence of hydrogen and helium bound-free opacity guarantees that
the average energy and resultant heating per ionizing photon increases
with increasing radius.  The combined effects are expected to produce
a gradient $dT_e/dr > 0$ in high metallicity \h2 regions\footnote{At
very low $Z$ the gradient in heating stemming from bound-free opacity is
overwhelmed by the role of Ly$\alpha$ cooling in determining the
equilibrium $T_e$, which consequently decreases with $r$.}.  Infrared
transitions also play an important role in cooling the
lower-ionization, outer portions of a nebula, particularly by emission
of [Si\,{\smc II}] 34.8 $\mu$m, lines of [Fe\,{\smc II}] (e.g., see
\cite{Nussbaumer88} 1988), and [C\,{\smc II}] 158 $\mu$m.

Grains and associated depletion may have an important role in
modifying the thermal structure of GEHRs beyond a simple global
increase in $T_e$, when compared with the dust-free case.  Examination
of the elemental depletion factors given in Table 1 indicates that
depletion has a much stronger influence on elements dominating
fine-structure cooling in the low-ionization, outer portions of a
nebula (Si, Fe, C) than on those important in the inner nebula (O, S,
N).  Depletion will consequently tend to steepen temperature gradients
from the undepleted case, by preferentially elevating the temperature
of the outer nebula.  This steepening will be mitigated, however, through
heating of the inner nebula by grains formed from the depleted
elements.  As noted in \S\ref{general}, the rate of grain
heating will tend to scale locally with the flux of ionizing
radiation.  Radial dilution and attenuation of the radiation produced
by a central star or stars will thus lead to a radial falloff in grain
heating within a nebula, assuming a uniform distribution of dust.

These consequences can be seen quantitatively in the middle panels of
Figures 6 and 7, which show $T_e$ profiles for the calculations
described in \S\ref{Calculations}, for the case of $Z = $ Z$_\odot$.
The results of adding the effects of dust incrementally are
illustrated by the successive curves.  For comparison, the lower panels
show the characteristic ionization state of the gas as a function of radius,
as described by the ionization fraction of
O$^+$ and O$^{++}$.  Grain heating and cooling as a function of $r$ is
illustrated in the top panels.  Grain heating may exceed 70\% of the
total heating rate in the inner nebula, and decreases monotonically
with increasing radius through most of the \h2 region.  Near $R_s$,
the fractional heating may again become large; in this partially
ionized zone, hydrogen bound-free heating is rapidly diminishing, but
photons with energy $8 \simlt h\nu < 13.6$ eV continue to produce
significant grain photoelectrons.  Fractional cooling by dust is lower than
fractional heating at all radii, with the possible exception of the
outermost portions of the Str\"omgren sphere.  These results show a
general similarity to calculations with a blister geometry described
by \cite{Baldwin91} (1991).

The radial dependences of depletion and grain heating propagate into
the predicted fluxes of optical lines, as can be seen by a review of
Figures 1 and 2.  For both choices of $T_\star$, emission in \O2, \s2,
and \n2  (which represent relatively weakly ionized species) show a
strong response to depletion, while the response in \o3 is less
dramatic.  In contrast, \o3 shows a much greater sensitivity to grain
heating than do \O2, \s2, or \n2, for the nebula powered by the cooler
star.  The differences in the behavior of
\o3 between the two choices of $T_\star$ reflect the relative extent
of the O$^{++}$ zone (cf. Figs. 6 and 7).  The cooler star produces a
more concentrated region of O$^{++}$, which is consequently more
strongly affected by grain heating; the O$^{++}$ zone produced by the
hotter star overlaps more broadly with zones of low-ionization
species, introducing a greater sensitivity to depletion.

Grain heating might be expected to produce very
strong enhancements of high-ionization lines that form close to the
ionizing stars.  In the present calculations, however, an easily
observable signature of this effect is not predicted, because such lines
are relatively weak, even when dust heating is taken into account.
A tradeoff results in that a strong dust signature may be more
apparent when $T_\star$ is relatively low, due to the fact that the
emitting volume of a high-ionization species is then highly
concentrated to the region of strong heating by grains; but when this
volume is small, emission is likely to be weak except for a tracer
with high abundance, such as oxygen.  An illustration is provided in
the case of $T_\star = 38,000$ K by
\Ne3 $\lambda$3869/H$\beta$, which increases by a factor of $\sim
2$ at $Z = $ Z$_\odot$ and by an order of magnitude at $Z =
2$Z$_\odot$ when grain heating and cooling is introduced to the
calculation; but invariably the \Ne3 line flux is less than 1\% of
that of H$\beta$.  If $T_\star$ is as high as 45,000 K, the \Ne3 line
becomes more readily observable (\Ne3/H$\beta = 0.1 - 0.2$), but shows
only a weak sensitivity to dust thermal effects since the Ne$^{++}$
volume extends over a larger fraction of the nebula, including regions
where grain photoelectric heating is relatively unimportant.

Signatures of radially dependent grain heating may be more readily
detected in planetary nebulae, which can span a broader range of
ionization than is typically found in \h2 regions.  \cite{Keyes90}
(1990) performed a detailed analysis of the spectrum of the planetary
nebula NGC~7027, and noted a persistent difficulty in accounting for
the observed (strong) emission in \ne5 $\lambda\lambda$3346, 3426
after adjusting model nebular parameters so as to reproduce correctly
the strength of He\,{\smc II} lines.  Since the He\,{\smc II} features are
recombination lines while \ne5 is collisionally excited, the
unexpected strength of the latter emission might be explained by a
thermal mechanism such as grain heating in the inner nebula, if a
fraction of the dust survives in this environment.  Test calculations
with model parameters similar to those employed by Keyes et al. are
consistent with this interpretation.  Detailed study of other sources
is needed in order to ascertain whether grain heating is often an
important influence on the emergent spectrum of planetary nebulae.

Abundance analyses based on \h2 region spectra are potentially
sensitive to the details of nebular temperature gradients.  The
derivation of an ionic abundance from a collisionally excited emission
line requires the use of a $T_e$ value appropriate to that ion.
Nonuniform temperature structure necessitates the use of different
characteristic temperatures for different ions; in addition,
temperature-sensitive (auroral/nebular) line ratios of a single ion
can differ from ion-weighted temperatures that are physically
appropriate for extracting abundance information from line strengths.
Radial gradients are a specific case of temperature fluctuations that
contribute lingering uncertainty in the calibration of abundances
based on nebular studies (e.g., \cite{Peimbert67} 1967;
\cite{Peimbert93} 1993).  Detailed aspects of these problems have been
reviewed by \cite{Garnett92} (1992), who derived theoretical relations
between ion-weighted temperatures for various species, as well as
between line-ratio and ion-weighted temperatures.  Garnett's models do
not include the consequences of grains, although he notes their
possible significance for perturbing nebular temperature structure.

The behavior of ion-weighted temperatures in the present calculations
suggest some possible modifications to previous theoretical work.
Garnett's models included depletion of Si (which he excluded
completely) and Fe, and our calculations with only depletion are
consistent with his results.  In the preceding discussion, it was noted
that depletion selectively boosts $T_e$ in the outer nebula, while
grain heating is most important in the inner nebula.  As a result,
models that include depletion but neglect the thermal effects of the
corresponding grain population may overestimate the steepness of the
temperature gradient in an enriched nebula.  These expectations are
borne out in part in the present calculations, as can be seen by
examination of Figures 8 and 9, which plot the ion-weighted
temperatures $T({\rm O}^+)$ and $T({\rm S}^{++})$ as functions of
$T({\rm O}^{++})$.  These temperatures are calculated according to
equation (1) in \cite{Garnett92} (1992); the plotted points span
$Z = 0.1 - 3$ Z$_\odot$. Also shown in these figures are the trends
found by Garnett (and confirmed by us) for a nebula with depletion
alone, i.e., his equations (2) and (3), as well as diagonal lines of
uniform $T_e$.

At high $Z$, the calculations with depletion and grain effects show a
more nearly isothermal behavior than predictions restricted to
depletion, for the case of the cooler $T_\star$.  For comparison, the
predictions for a dust-free and undepleted nebula are also shown;
interestingly, these predictions are in better agreement with the
calculations with full dust effects than with those taking only
depletion into account.  This behavior reflects a rough tendency for
depletion and grain heating to balance each other in terms of their
net effect on $dT_e/dr$.  For the hotter $T_\star$, $T({\rm S}^{++})$
at a given $T({\rm O}^{++})$ for the dusty nebula is still somewhat
overpredicted by the relation with depletion only, while $T({\rm
O}^{+})$ versus $T({\rm O}^{++})$ is consistent with the
depletion-only results.  For $T_\star = 45,000$ K, S$^{++}$ occupies
an intermediate zone in radius that experiences a smaller relative
increase in $T_e$ than either of the inner or outer regions weighted
respectively by O$^{++}$ or O$^+$; the overall gradient in $T_e$
remains positive for the most part, however, such that the relation
between $T({\rm O}^{++})$ and $T({\rm O}^+)$ does not change
appreciably from the dust-free or depletion-only case.  Calculations
with $n_e$ increased to 100 and 1000 cm$^{-3}$ produce a general
increase in nebular temperatures, due to partial collisional
suppression of fine-structure cooling, but the ion-weighted temperature
behavior is qualitatively the same as seen in Figures 8 and 9.  These
results also generally show a weak sensitivity to $U$.  Increasing
$\epsilon$ such that log $U = -2.5$ causes the {\it dust-free}
calculations for $T_\star = 38,000$ K to approach Garnett's relations,
since the inner zone of high-ionization that cools efficiently
expands, resulting in a steeper $T_e$ gradient; the results for the
dusty nebula are qualitatively unchanged, however.  We find reasonable
agreement in the present calculations with relations listed by
\cite{Garnett92} for other ion-weighted temperatures [i.e., $T({\rm
Ar}^{++}) = T({\rm S}^{++})$; $T({\rm N}^+) = T({\rm O}^+)$; $T({\rm
N}^{++}) = T({\rm O}^{++})$; $T({\rm Ne}^{++}) \approx T({\rm
O}^{++})$; $T({\rm S}^+)
\approx T({\rm O}^+)$].

If enriched GEHRs contain significant quantities of dust, the present
results suggest that these nebulae may be treated as isothermal systems
in terms of the relative behavior of ion-weighted temperatures, to
good approximation.  While this statement is strictly true only for
the calculations presented here with the cooler $T_\star$, empirical
evidence suggests that the characteristic $T_\star$ of a nebula
decreases with increasing metallicity, such that a relatively soft
radiation field is, in fact, appropriate for enriched systems with
$T_e \simlt 7000$ K.  If dust is efficiently destroyed in
\h2 regions, an isothermal treatment may still be appropriate,
since the effects of depletion as well as grain heating would be
vitiated\footnote{A relevant technical point is that Cloudy employs a
modified on-the-spot (OTS) approximation for transfer of the hydrogen
and helium diffuse radiation field.  In some cases OTS can lead to an
estimated temperature gradient that is too steep compared to an
outward-only approximation (which was employed by \cite{Garnett92}
1992) or a rigorous treatment of radiative transfer (i.e., $dT_e/dr$
is too large; see, for example, discussion in \cite{Sarazin77} 1977;
\cite{McCall85} 1985; \cite{Ferland95} 1995), which suggests that our
results are conservative in that we are unlikely to predict a
(positive) $dT_e/dr$ that is too shallow.}.

We note that nebular models with depletion alone may
nonetheless produce accurate predictions if grains are preferentially
destroyed at small radii.  In this situation the effects of grain
heating that predominate at small $r$ would be removed, while
depletion effects that selectively influence the outer nebula would
still be relevant.  Evidence for reduced depletion in the inner,
high-ionization regions of planetary nebulae has been discussed
recently by \cite{Kingdon95} (1995).

Modifications to relations between line-ratio-inferred $T_e$ and
ion-weighted temperatures will also result from dust effects.  In the
present calculations the largest discrepancies between the two
temperature values occur at ion-weighted values of $T_e \simlt 6000$
K; in this domain, auroral transitions employed for direct estimation
of $T_e$ are generally too weak to be observable.  Additional
perturbations in temperature estimates from line ratios (and also
ion-weighted $T_e$ values) may result from more complicated spatial
distributions of dust, as noted by Peimbert et al. (1993).

As a final cautionary note, we remind the reader that the present
calculations (and many others in the literature) assume a constant
density, independent of $r$.  While a detailed inquiry into the
effects of nonuniform density is beyond the scope of the present work,
line fluxes and temperature structure can certainly deviate in
important ways from the predictions given here if large gradients in
density exist within a nebula.  Observations suggest that $n_e$ is
often a decreasing function of $r$ in GEHRs (e.g., \cite{Kennicutt84}
1984).  For \h2 regions with gradients of this form, the temperature
of the inner nebula may be boosted and positive gradients in $T_e$
again diminished if $n_e$ at small radii is sufficient to partially
suppress cooling via infrared fine-structure transitions with low
critical densities.  In general terms, however, the physical effects
of the generic grains discussed in this paper are independent of the
density structure of the nebula in which the dust resides.


\section{Conclusions}

Embedded dust can modify the optical spectrum of GEHRs, through a
combination of effects that mostly act to increase $T_e$.  The
largest perturbations are generally the result of depletion of
coolants from the gas phase, although grain opacity and heating may be
significant in some cases.  The net result is usually an increase in
optical forbidden-line strengths, particularly at high metallicity.
Dust may lead to nonnegligible modification of nebular behavior
without introducing strong signatures of reddening or extinction.

The influence of grains provides a simple way of explaining the
observational lack of \h2 regions emitting a pure Balmer-line
spectrum at optical wavelengths, which is expected on theoretical
grounds for dust-free nebulae with metallicity only slightly greater than
Z$_\odot$.  At high $Z$, predicted enhancements of \s2 and \n2 in
particular show a resemblance to the empirical behavior of \h2 regions
in galaxy nuclei.  If these environments are particularly dusty,
due perhaps to characteristically elevated densities, the influence
of dust provides an alternative mechanism to weak active nuclei
for explaining the anomalous line ratios of these sources.

Depletion and heating by photoelectrons emitted by grains in \h2
regions may introduce significant modifications to nebular temperature
structure.  Depletion selectively influences elements that are
important for cooling the outer portions of nebulae, while grain
heating is most effective at small radii.  Modifications to
temperature gradients within GEHRs by dust may necessitate changes in
assumptions employed in nebular abundance analyses, specifically in
the relationships between ion-weighted temperatures for different
species.  For chemically enriched GEHRs, the present results suggest
that these nebulae can be treated as nearly isothermal systems.

Inclusion of grain physics is unlikely to result in large changes of
gas-phase heavy-element abundances derived for low-$Z$ objects and
GEHRs ionized by relatively hot stars.  In general, dust-related
phenomena will contribute an uncertainty of $\sim 0.2$ dex or less in
gas-phase oxygen abundances derived from the $R_{23}$ parameter.
Modifications to nebular temperature structure by dust and associated
depletion may be significant in \h2 regions described by high $Z$ and
low $T_\ast$, and should be considered when deriving relative abundances
for heavy elements.

\acknowledgements

This work was supported by NASA through grant number HF-1052.01-93A
from the Space Telescope Science Institute, which is operated by the
Association of Universities for Research in Astronomy, Inc., under
NASA contract NAS5-26555.  Additional support was provided by the
National Science Foundation through grants AST-9019150 and
AST-9421145.  We thank Gary Ferland for conversations that stimulated
this work, as well as for access to Cloudy.  Informative discussions
with John Black, Don Garnett, Sally Oey, and Evan Skillman are
gratefully acknowledged.  We also thank the anonymous referee for
helpful comments on the manuscript.


\begin{table*}
\begin{center}
\begin{tabular}{lcr}
 & {TABLE 1} & \\
 & ~~~~~~~~~~~~ & \\
 & ~~~~~~~~~~~~~~~~~~~~~~~~~{\smc Relative Abundances$^{\rm a}$ at $Z = {\rm
Z}_\odot$}~~~~~~~~~~~~~~~~~~~~~~~~~ & \\
 & ~~~~~~~~~~~~ & \\
\end{tabular}
\begin{tabular}{lrr}
\tableline
\tableline
Element & Total & Depleted \\
\tableline
He &  10.99 &   10.99~~~\\
C  &   8.56 &    8.22~~~\\
N  &   8.05 &    7.95~~~\\
O  &   8.93 &    8.73~~~\\
Ne &   8.09 &    8.09~~~\\
Na &   6.30 &    5.50~~~\\
Mg &   7.58 &    7.15~~~\\
Al &   6.47 &    5.53~~~\\
Si &   7.55 &    6.55~~~\\
S  &   7.21 &    7.21~~~\\
Ar &   6.56 &    6.56~~~\\
Ca &   6.36 &    3.66~~~\\
Fe &   7.67 &    5.97~~~\\
Ni &   6.25 &    4.26~~~\\
\tableline
\end{tabular}
\begin{tabular}{c}
{}~~~~~~~~~~~~ \\
$^{\rm a}$Abundances are logarithmic + 12, by number relative to hydrogen. \\
\end{tabular}
\end{center}
\end{table*}

\clearpage

{}~~~~~~~~~~~~~~~~~~~~~~~~~~~~~~~~~~~~~~~~~~~{\bf Figure Captions}

Fig. 1 --- Line ratios plotted as a function of total metallicity, with
dust effects added incrementally, for $T_\star = 38,000$ K.  {\it
Solid curves:} no depletion or grains; {\it dotted:} depletion only;
{\it short dashed:} depletion + grain opacity; {\it long dashed:}
depletion, with grain opacity, heating, and cooling.  The lines
\O2 $\equiv \lambda\lambda3726, 3729$; \o3 $\equiv \lambda\lambda4959, 5007$;
\s2 $\equiv \lambda\lambda6716, 6731$; \n2 $\equiv \lambda\lambda6548, 6583$.
Approximate lower bounds to observed ratios in \h2 regions are indicated
by the dotted horizontal lines.

Fig. 2 --- Same as Figure 1, for $T_\star = 45,000$ K.

Fig. 3 --- Loci of line ratios resulting from variation in $Z$ ($=
0.1 - 5$ Z$_\odot$), for dust-free (solid curves) and dusty (dotted
curves) nebulae.  Values of $Z$ are indicated by the corresponding
filled points.  Crosses represent measured, dereddened values for
GEHRs taken from \cite{McCall85} (1985).

Fig. 4 --- Same as Figure 3, with axes chosen for comparison with
results for nuclear and disk GEHRs studied by \cite{KKB89} (1989).
A general discussion of the loci of emission-line nuclei and other
objects in line-ratio diagrams can be found in \cite{veilleux87}
(1987).

Fig. 5 --- Values of $R_{23}$ as a function of gas-phase oxygen
abundance.  Results are shown for nebular cases that are dust-free
(solid curves) and dusty (dotted curves), with oxygen abundance
normalized to log (O/H)$_\odot = -3.07$ by number.  The upper two
curves assume $T_\star = 45,000$ K, while the lower pair assumes
$T_\star = 38,000$ K.  The abundances of other heavy elements scale
with oxygen as described in \S 2.

Fig. 6 --- Radial behavior of thermal and ionization properties for
$T_\star = 38,000$ K and $Z =$ Z$_\odot$.  Central panel: $T_e$ as a
function of radius, with depletion and grain effects added
incrementally.  Top panel: heating and cooling contributed by grains
as a fraction of the total.  Bottom panel: Ionization fraction of
O$^+$ and O$^{++}$; solid curves represent a dust-free nebula, while
the dashed curves show the result after adding full dust physics.

Fig. 7 --- Same as Figure 6, for $T_\star = 45,000$ K.

Fig. 8 --- $T({\rm O}^+)$ versus $T({\rm O}^{++})$ for the dust-free
and dusty calculations, where the temperatures represent ion-weighted
averages.  The dashed line represents the theoretical relation
obtained by \cite{Garnett92} (1992) for depleted nebulae, while the
dotted line represents the isothermal case.

Fig. 9 --- $T({\rm S}^{++})$ versus $T({\rm O}^{++})$ for the
dust-free and dusty calculations.  The dashed line represents the
theoretical relation obtained by \cite{Garnett92} (1992) for depleted
nebulae, while the dotted line represents the isothermal case.

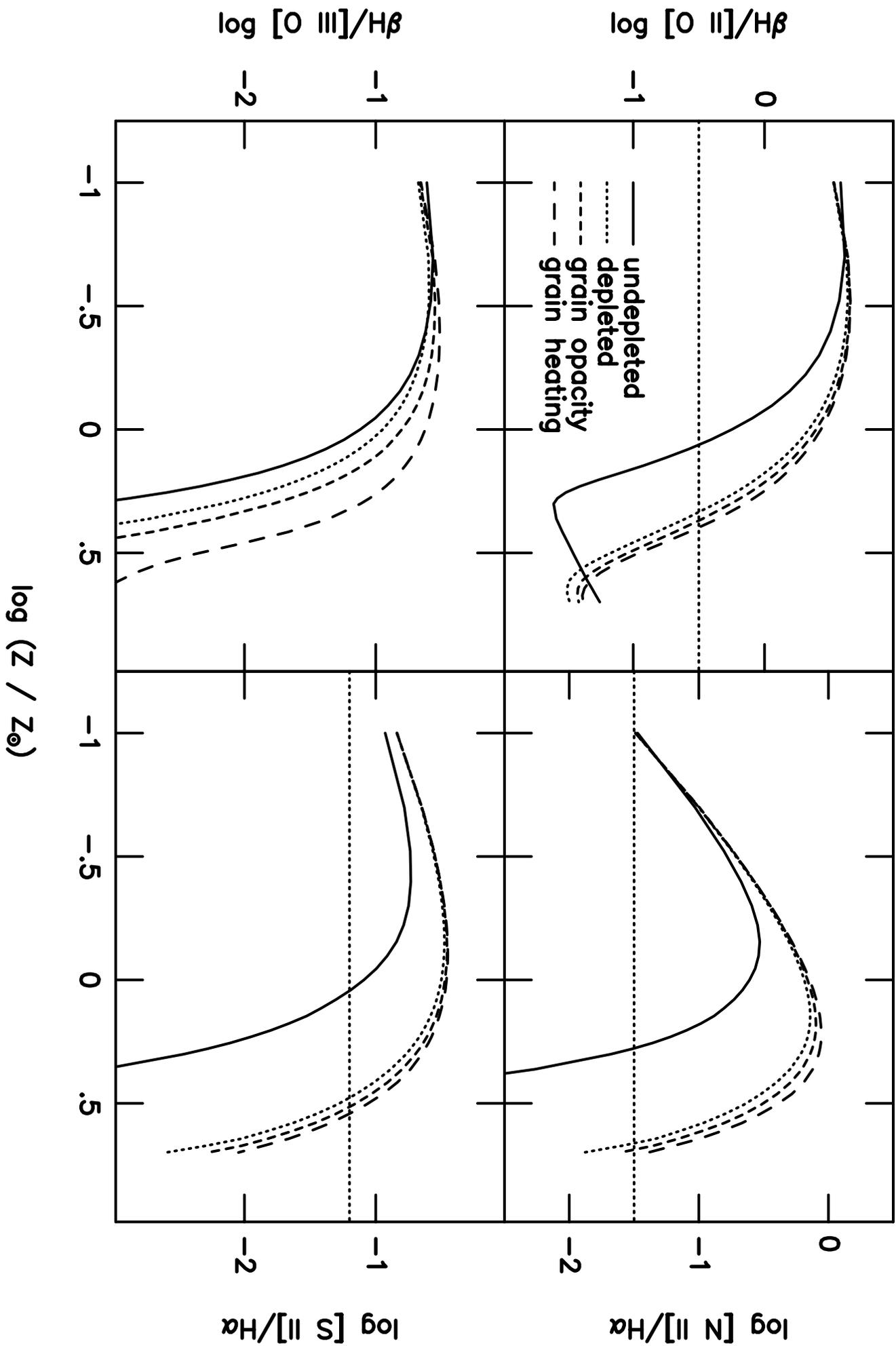
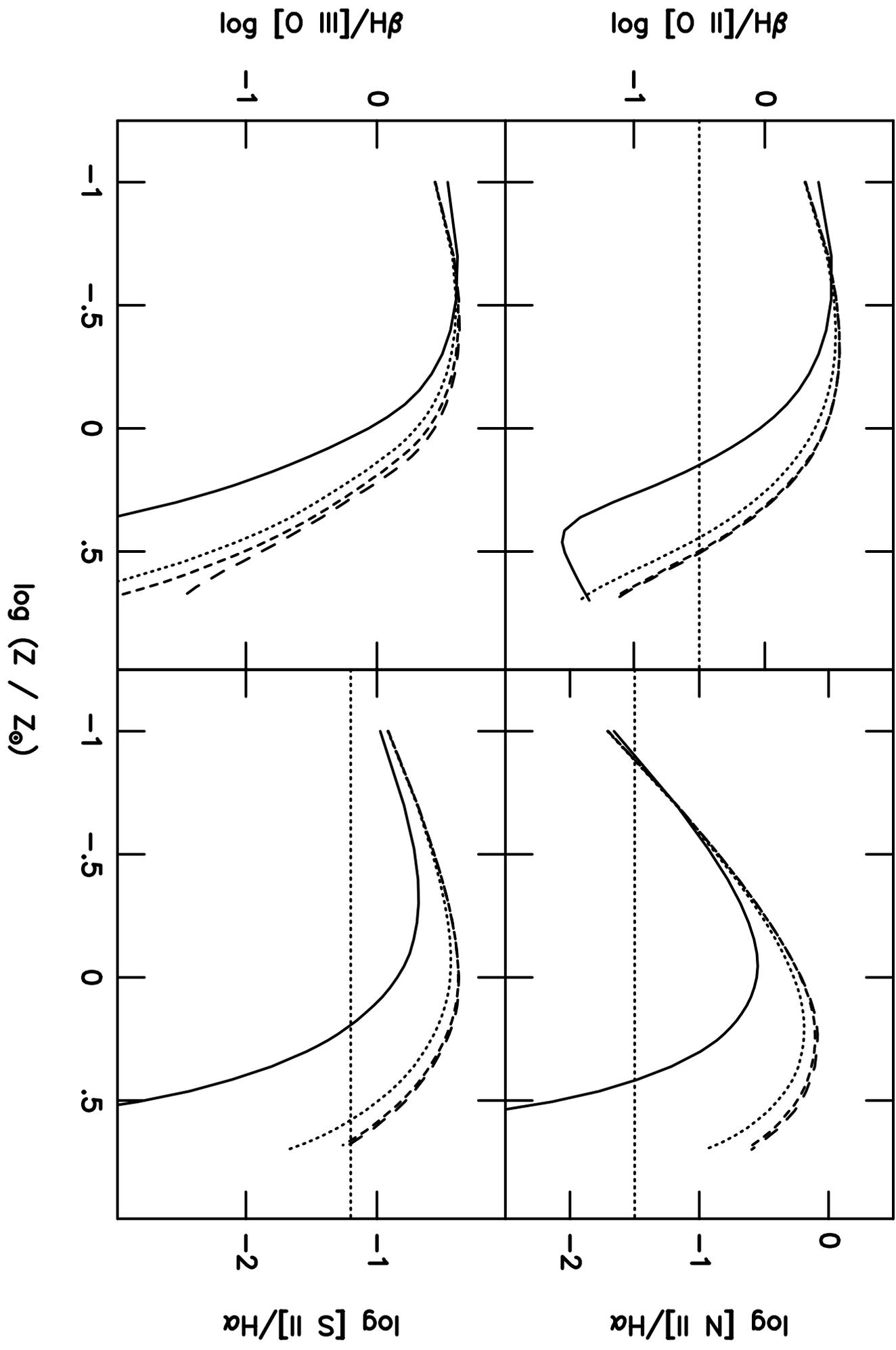
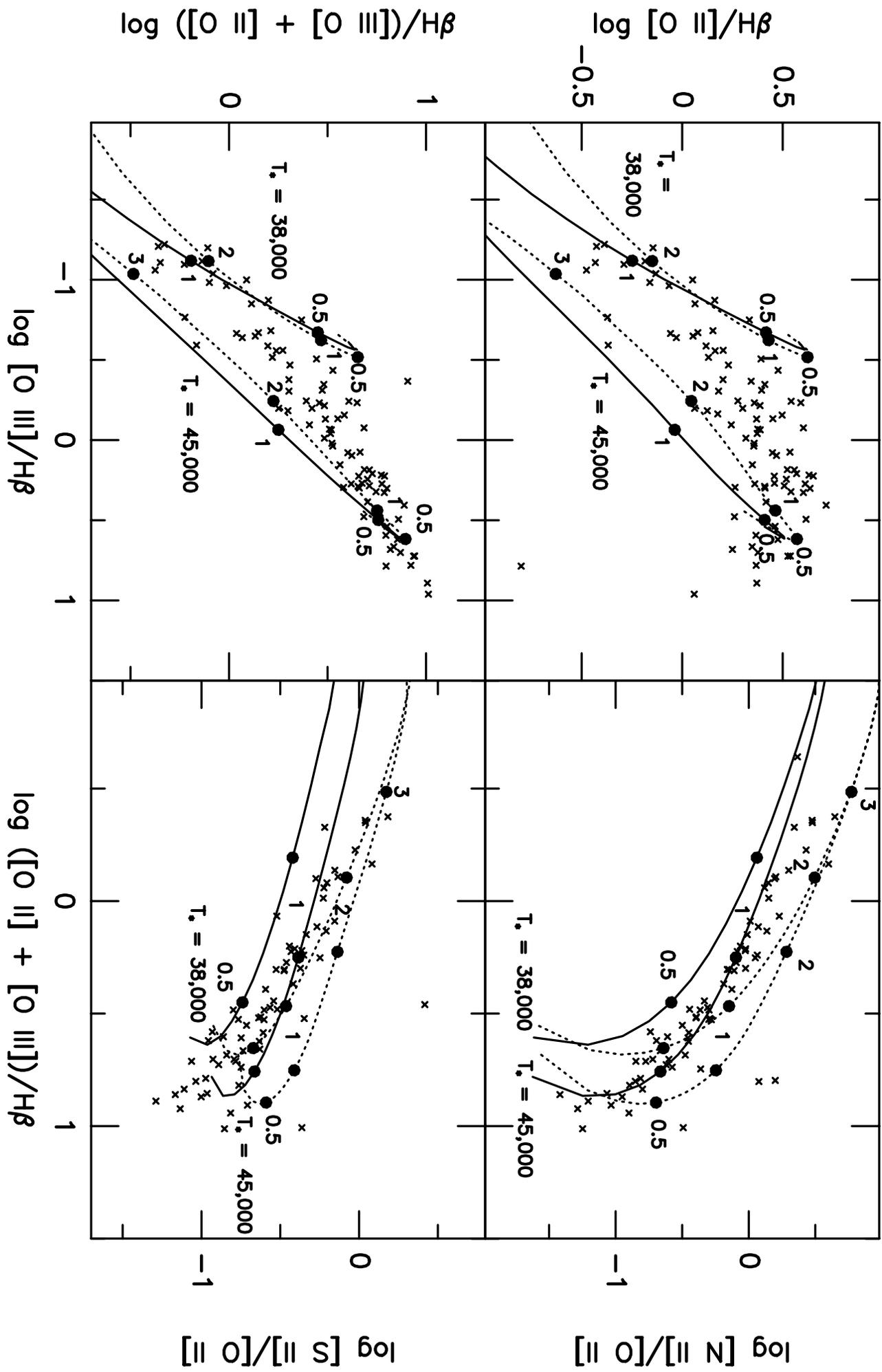
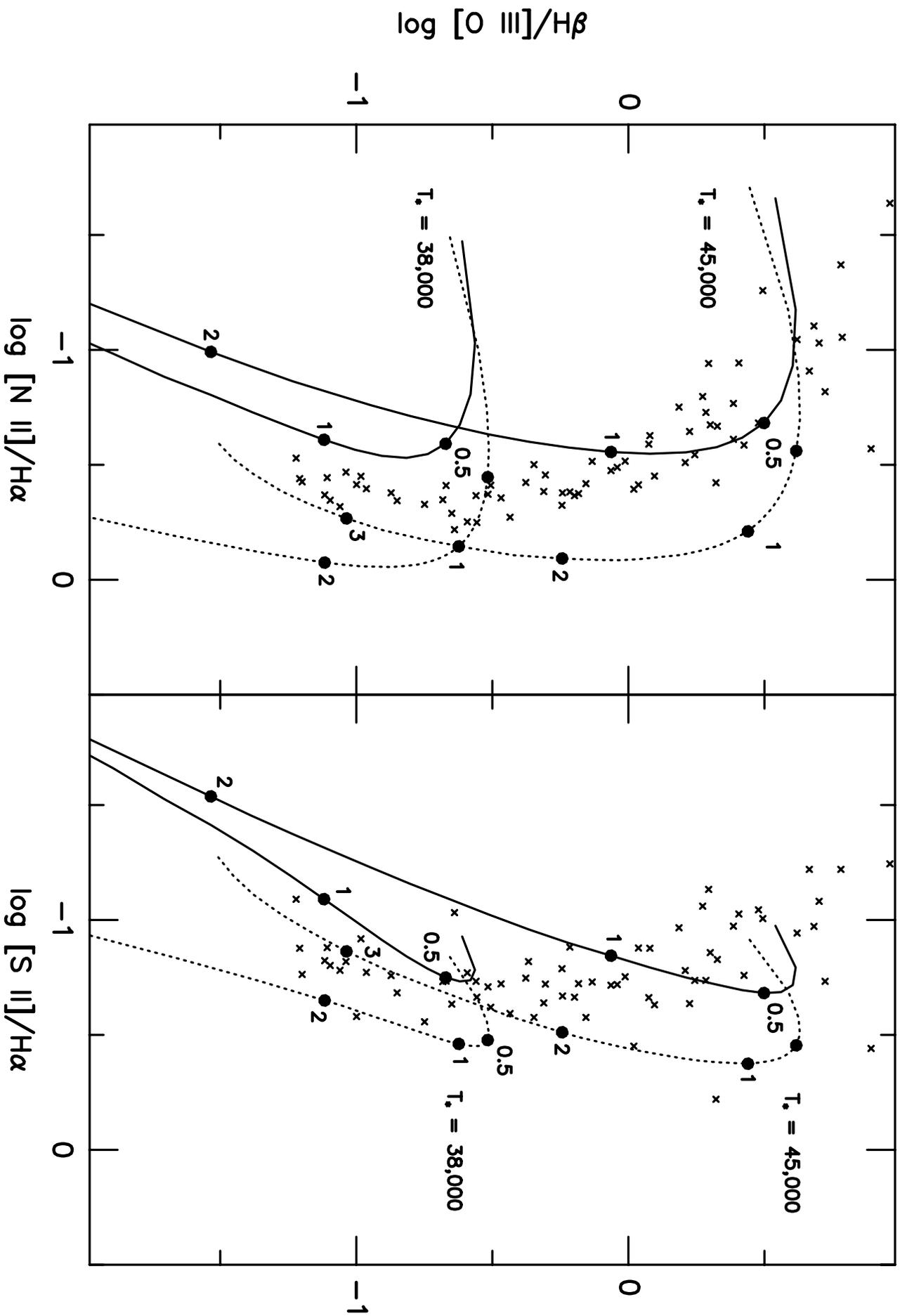
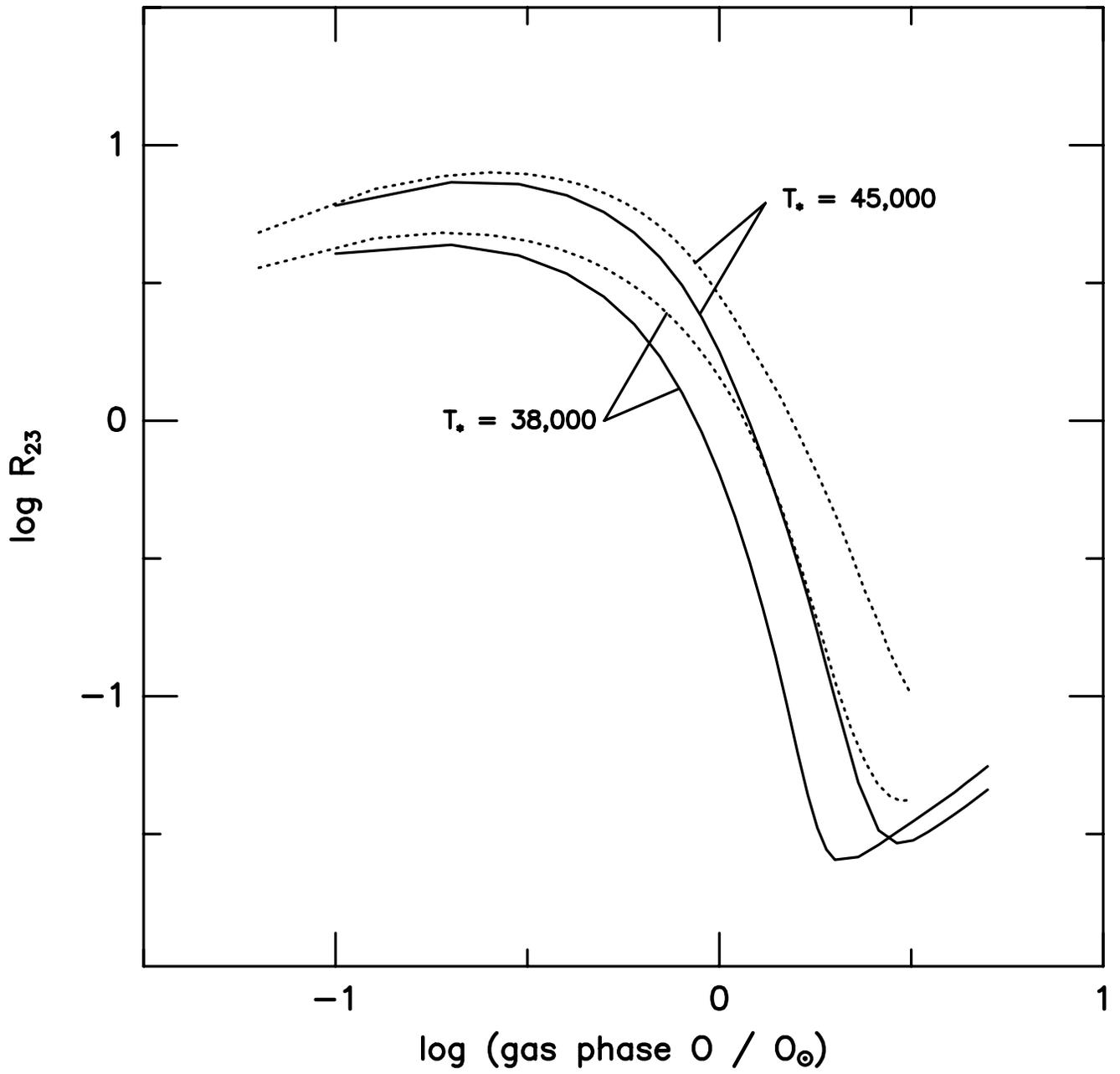
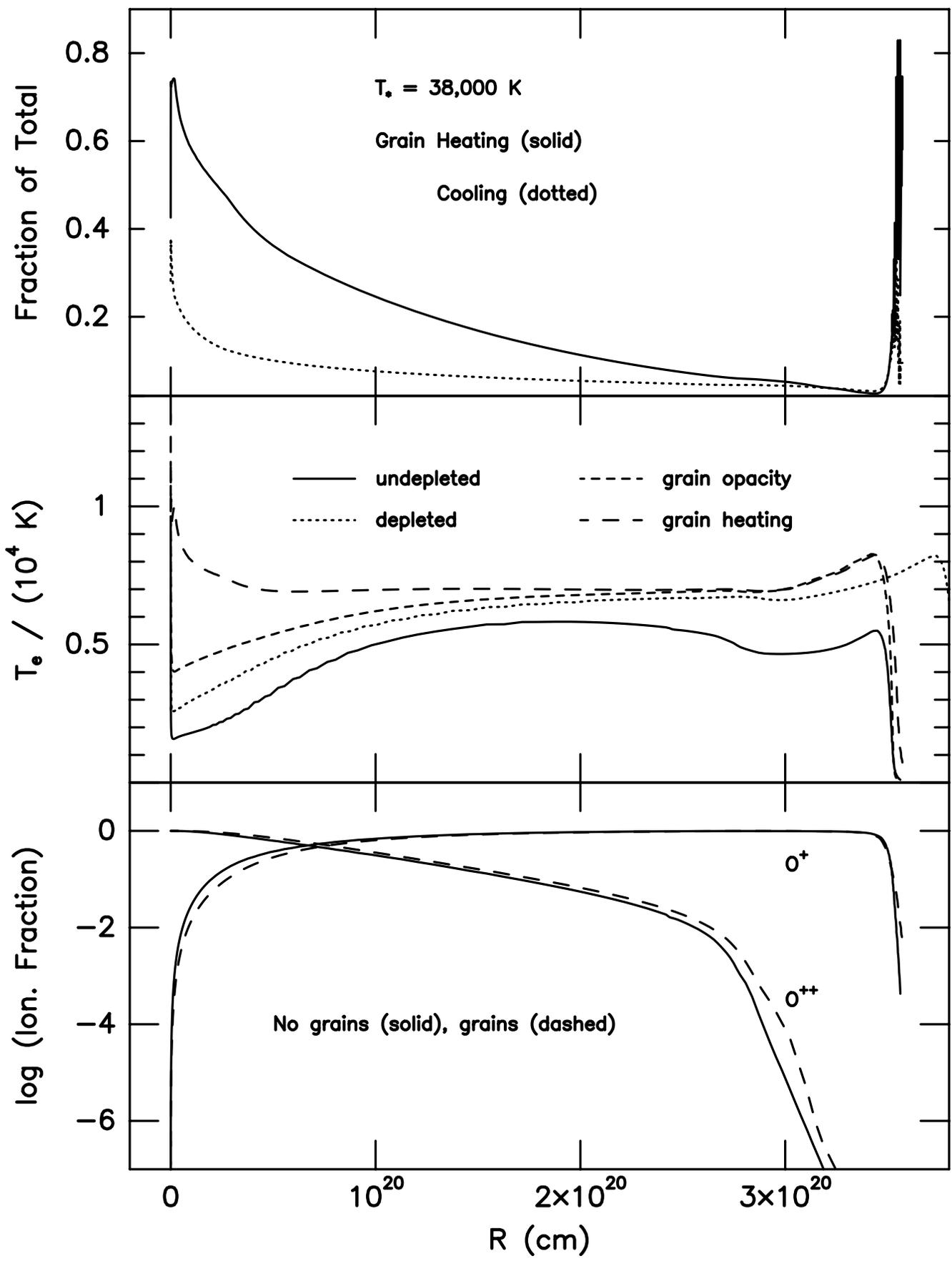
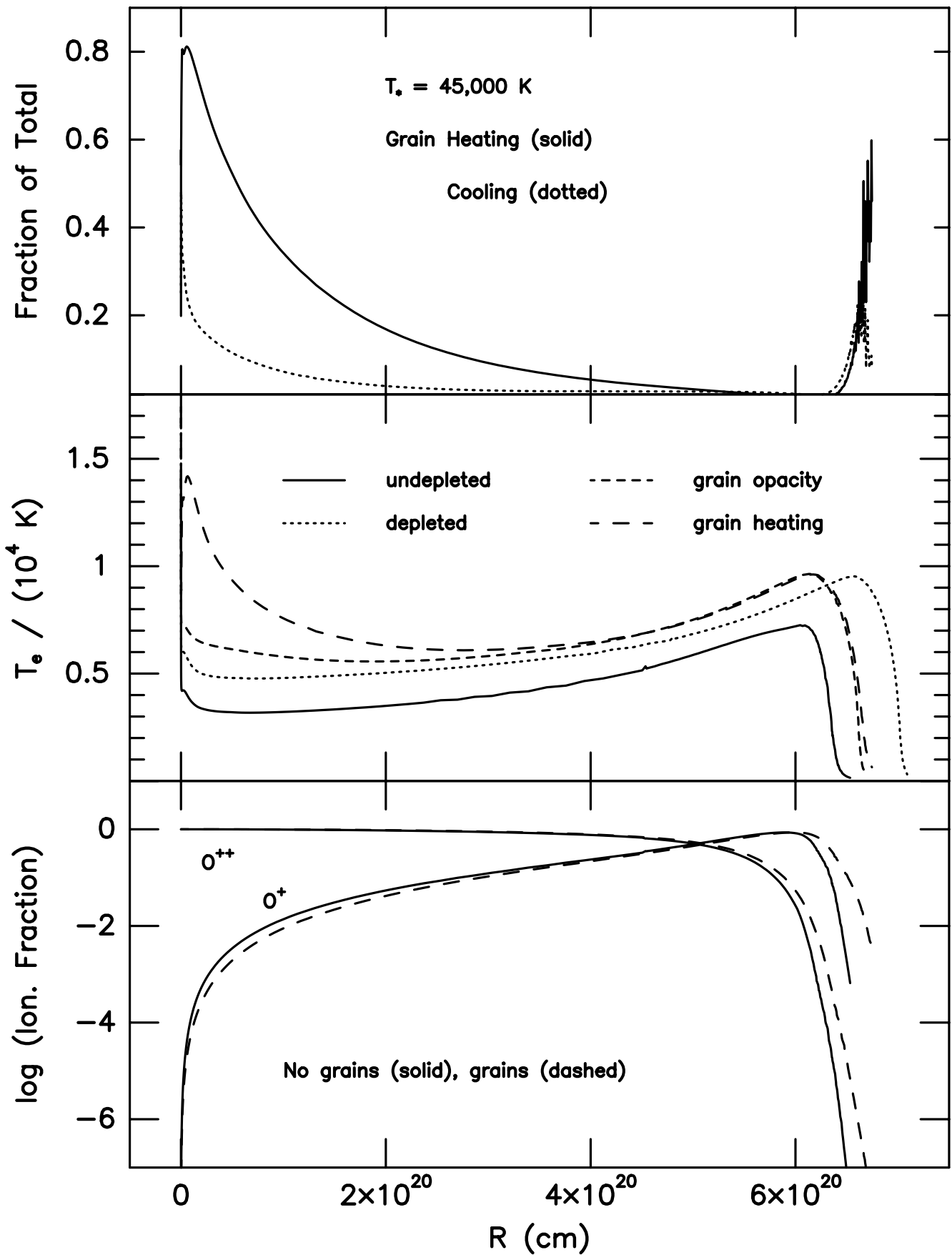
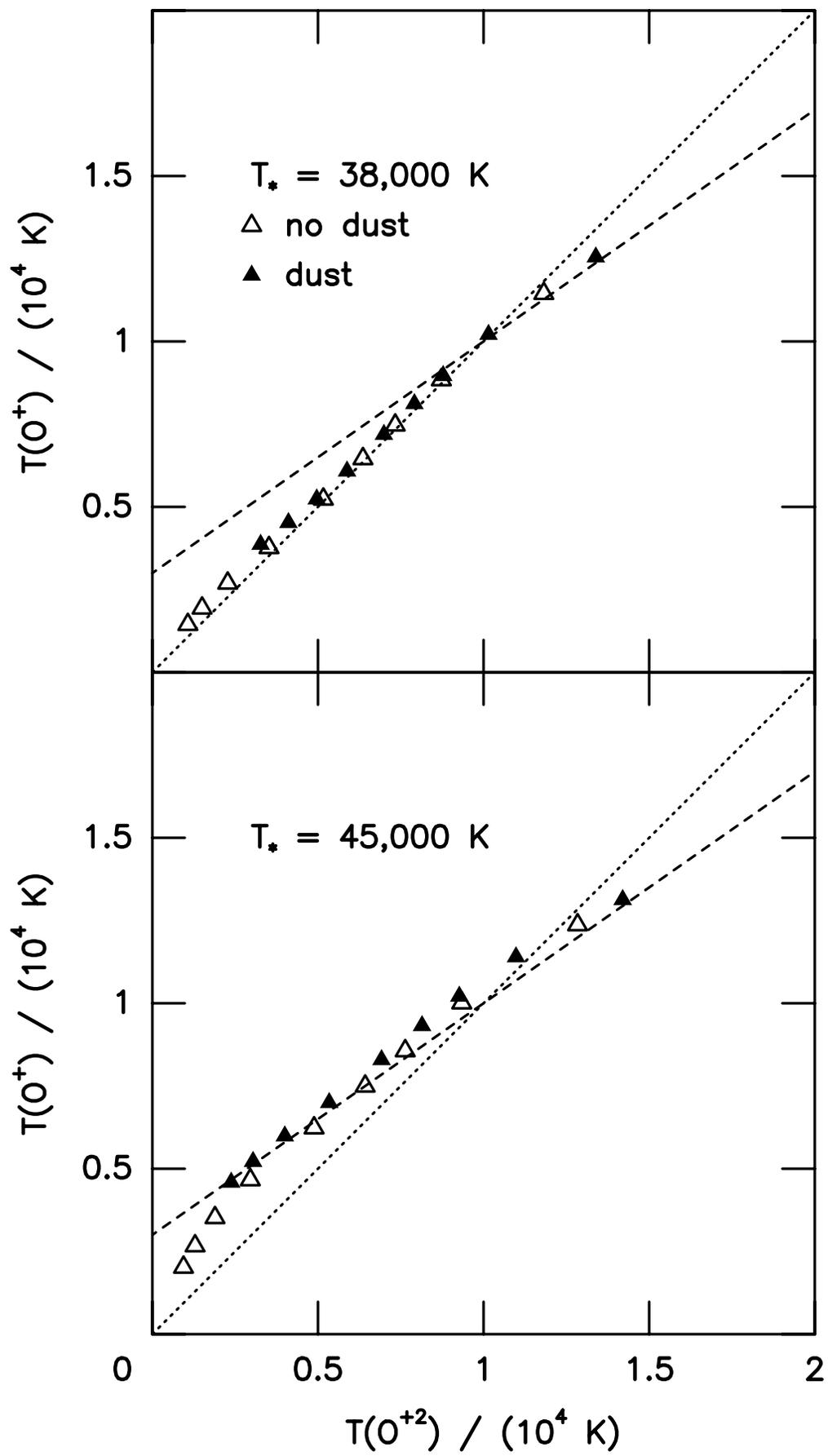
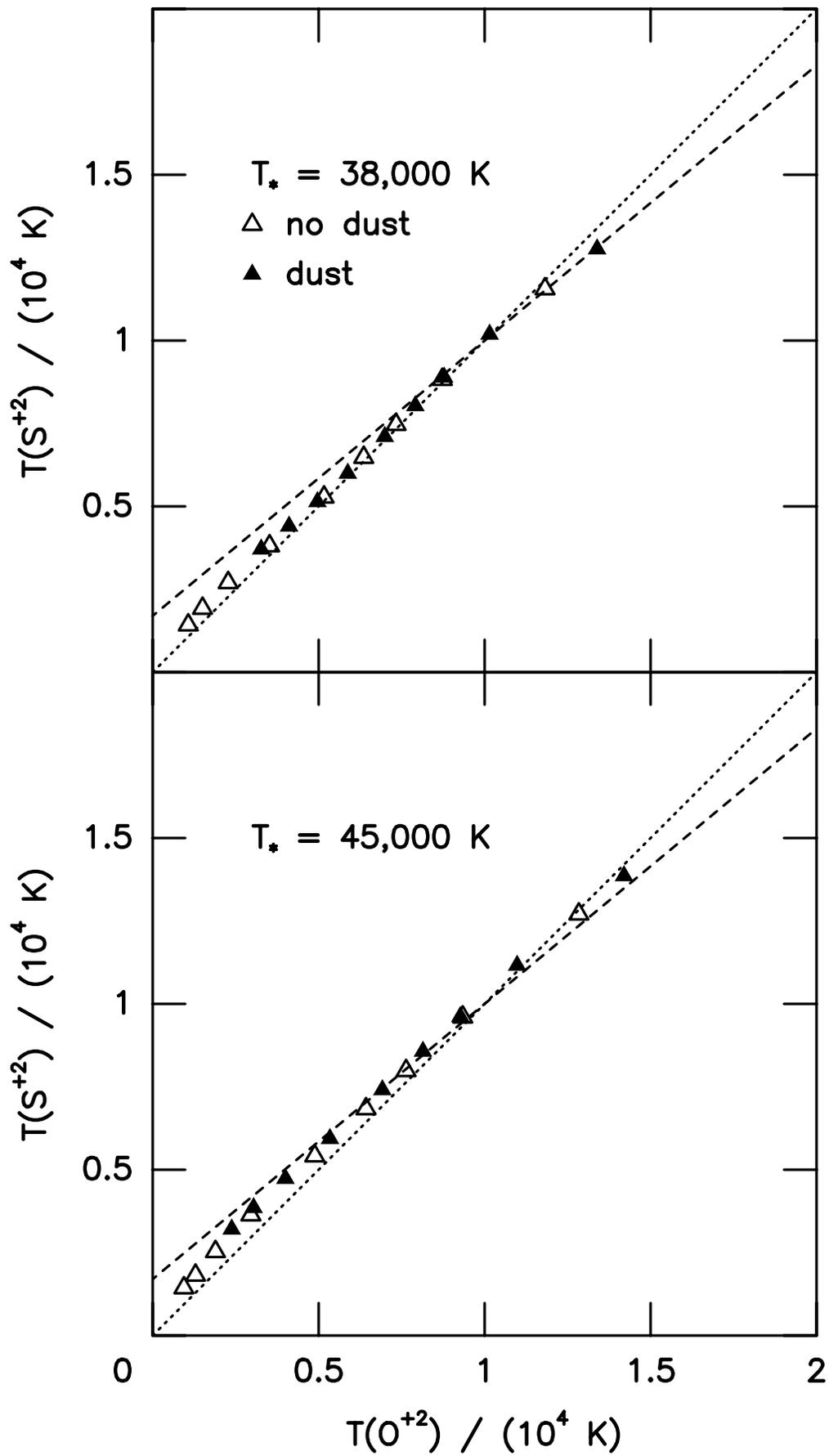
\end{document}